\documentclass[aps,pre,twocolumn,showpacs,preprintnumbers,amsmath,amssymb,10pt,a4paper]{revtex4}

\usepackage{graphicx}
\usepackage{dcolumn}
\usepackage{bm}
\usepackage{subfigure}
\usepackage{amsmath}
\usepackage{amsfonts}
\usepackage{multirow}
\newcommand{\be}{\begin{equation}}
\newcommand{\ee}{\end{equation}}
\newcommand{\bd}{\begin{displaymath}}
\newcommand{\ed}{\end{displaymath}}
\newcommand{\BE}{\begin{eqnarray}}
\newcommand{\EE}{\end{eqnarray}}

\newcommand{\bh}{\ensuremath{\mathbf{h}}}

\newcommand{\bn}{\ensuremath{\mathbf{n}}}

\newcommand{\shifta}{\widehat{E}_1} 
\newcommand{\shiftb}{\widehat{E}_2} %
\newcommand{\shifti}{\widehat{E}_i}
\newcommand{\pd}[2]{\frac{\partial #1}{\partial #2}}

\begin{document}

\preprint{}
\title{Evolutionary dynamics, intrinsic noise and cycles of co-operation}

\author{Alex J. Bladon}
\email{alex.bladon@postgrad.manchester.ac.uk}
\author{Tobias Galla}
\email{tobias.galla@manchester.ac.uk}
\author{Alan J. McKane}
\email{alan.mckane@manchester.ac.uk}

\affiliation{Theoretical Physics, School of Physics and Astronomy, University 
of Manchester, Manchester M13 9PL, United Kingdom}

\date{\today}

\begin{abstract}
We use analytical techniques based on an expansion in the inverse system 
size to study the stochastic evolutionary dynamics of finite populations of 
players interacting in a repeated prisoner's dilemma game. We show that a 
mechanism of amplification of demographic noise can give rise to coherent 
oscillations in parameter regimes where deterministic descriptions converge 
to fixed points with complex eigenvalues. These quasi-cycles between 
co-operation and defection have previously been observed in computer 
simulations; here we provide a systematic and comprehensive analytical 
characterization of their properties. We are able to predict their power 
spectra as a function of the mutation rate and other model parameters, and 
to compare the relative magnitude of the cycles induced by different types 
of underlying microscopic dynamics. We also extend our analysis to the 
iterated prisoner's dilemma game with a win-stay lose-shift strategy, 
appropriate in situations where players are subject to errors of the 
trembling-hand type.
\end{abstract}


\maketitle


\section{Introduction}
\label{sec:intro}

Traditionally, modellers in biology and related disciplines use deterministic 
ordinary or partial differential equations to capture the quantitative 
behavior of dynamical systems in those fields. Such an approach is valid and 
accurate only if stochastic effects induced by external or intrinsic 
fluctuations can be neglected. External noise might result from 
environmental factors or as an attempt to include the effects of numerous,
but weak, external effects. Intrinsic fluctuations arise from the dynamics 
of the system itself. One of the most common sources of such stochasticity 
in biology is discretization noise in systems composed of a {\em finite} 
number of interacting individuals. While deterministic descriptions can be 
derived, and shown to be exact in the limit of infinite system size, finite 
systems retain an intrinsic randomness, sometimes referred to as demographic 
noise~\cite{Nisbet1982}. Such fluctuations can invalidate 
conclusions based on the analysis of the deterministic dynamics, turning 
deterministic fixed points into stochastic quasi-cycles, inducing helical 
motion about limit cycles~\cite{Boland2009}, or giving rise to Turing patterns 
induced by intrinsic noise~\cite{Butler2009,Biancalani2009}. The existence of stochastic quasi-cycles has been known for a number of decades
in the context of predator-prey-like systems, and methods have been devised to distinguish them from noisy limit cycles \cite{PinedaKrch2007} .

Only very recently have systematic methods, based on a system-size expansion of the master equation 
describing the microscopic stochastic processes, been devised to study them 
analytically~\cite{McKane2005}. These methods use an expansion in the inverse 
system size~\cite{Kampen2007}, and are now being applied to a number of 
fields in which quasi-cycles have been reported, including 
epidemiology~\cite{Alonso2007,Black2009,Rozhnova2009}, biochemical 
reactions~\cite{McKane2007}, gene regulation~\cite{Galla2009}, and more 
recently learning algorithms of interacting agents~\cite{Galla2009b}. The 
purpose of the present work is to apply these ideas to problems in 
evolutionary game theory, and to provide an analytical characterization of 
stochastic quasi-cycles found in computer simulations of populations of 
interacting players~\cite{Imhof2005}.

Evolutionary dynamics in this context is a mathematical framework 
describing co-evolving populations. It is the main tool-kit used in
attempts to reconcile the evolution of co-operation with Darwinian natural 
selection --- a problem which was listed as one of the $25$ most pressing 
scientific challenges in Science magazine in 2005~\cite{Pennisi2005}. The 
problem of how mutual co-operation is sustained in a population subject to 
selection pressure favoring selfish behavior is most commonly modeled using 
the prisoner's dilemma (PD) game~\cite{Axelrod1981,Nowak2006}. The PD is a 
classic game-theory problem in which two players have to simultaneously 
choose whether to co-operate or to defect. Although the payoff for mutual 
co-operation is higher than that for mutual defection, the payoff for 
defecting when the other player co-operates is higher still. Defection then 
forms the Nash equilibrium of the game, i.e. the outcome one may expect if 
the interacting players are fully rational. A number of experiments have been performed in behavioral game theory (examples are~\cite{Traulsen2010,Semmann2003}) and biological realizations of the PD include the study of competitive interaction among viruses, see e.g.~\cite{Turner1999}.
An extension of the basic PD game considers repeated interaction of a given 
pair of players. The space of available strategies then becomes too large to 
allow for an exhaustive analysis. Most studies therefore focus on a selected 
set of strategies, such as always defect (AllD), always co-operate (AllC), 
tit-for-tat (TFT) or win-stay lose-shift (WSLS). AllC players always 
co-operate in any iteration, and similarly AllD players always defect. TFT 
co-operates in the first round and then copies its opponent's previous move. 
This strategy emerged as the winner in a computer tournament run by Axelrod 
in 1981~\cite{Axelrod1981}. Since then TFT has been the subject of a large 
body of work~\cite{Axelrod1988,Nowak2004,Imhof2005,Imhof2007}. Even though 
TFT won a subsequent second competition as well, TFT is not perfect. In 
more realistic situations where players can make mistakes TFT can become 
locked into patterns of alternative co-operation and defection with another 
TFT player~\cite{Nowak1992b}. It is also vulnerable to invasion from 
co-operators via neutral drift. Nowak and Sigmund~\cite{Nowak1993} then 
proposed WSLS; this strategy has none of the above disadvantages. WSLS 
co-operates in the first round and then keeps playing the same action 
(co-operate or defect) if it receives a favorable payoff, and switches from 
one action to the other if it does not. It can resist neutral drift by 
co-operators and can correct mistakes, avoiding disadvantageous cycles. 
There is evidence to suggest that some animals employ these strategies, 
for example, in their behavior in the presence of 
predators~\cite{Milinski1987,Lombardo1985}.

Historically, the analysis of evolutionary dynamics has mostly been based 
on deterministic replicator dynamics~\cite{Taylor1978}, explicitly 
excluding stochastic effects. More recently, methods from statistical 
physics and the theory of stochastic processes have been used to study 
games in finite populations. In the absence of mutation, a finite 
population will always fix on a given strategy due to stochastic 
fluctuations. The resulting fixation probabilities and average fixation 
times can be calculated~\cite{Altrock2009,Traulsen2007,Altrock2009b}. 
Further quantities of interest are stationary distributions of the underlying 
stochastic processes~\cite{Traulsen2006,Imhof2007,Claussen2005,Fudenberg2006}, 
and the phenomenon of dynamic drift \cite{Claussen2007}. 

In the context of these studies of stochastic processes in game theory, cyclic 
behavior has been reported~\cite{Reichenbach2007,Reichenbach2006,Mobilia2009} 
in the rock-papers-scissors game, and in~\cite{Imhof2005}, where stochastic 
quasi-cycles between co-operation and defection have been observed in finite 
populations of agents playing the iterated PD. In the present work we will 
focus on the latter game, and provide an analytical theory which allows one 
to compute properties such as power spectra, or equivalently the correlation 
functions of these quasi-cycles, to a good approximation in the limit of 
large, but finite populations. Based on this analytical approach we are able 
to identify regions in parameter space where stochastic quasi-cycles would 
be expected to occur, and we compare the amplitude of cycles arising from 
different types of microscopic update dynamics. 

The paper is organized as follows. In Sec.~\ref{sec:IPD} we outline the 
iterated PD and define the different microscopic processes. We first focus 
on the case of only three pure strategies, AllC, AllD, and TFT. The 
deterministic analysis for this model is presented in 
Sec.~\ref{sec:IPDdeterministic} with a classification of the fixed points and 
an exploration of the parameter space. We move from a deterministic 
description to a stochastic formulation in Sec.~\ref{sec:IPDstoc} and consider 
effects arising in finite populations. In particular, we carry out a 
system-size expansion of the master equation allowing us to classify the 
periodic stochastic deviations from the deterministic limit. In 
Sec.~\ref{sec:IIPDE} we extend the analysis to include WSLS as a fourth 
strategy. Finally, in Sec.~\ref{sec:conclusion}, we summarize our findings 
and outline avenues of future research. 

\section{Model and definitions}
\label{sec:IPD}
\subsection{Iterated PD}
\label{sec:IPDmodel}

We will mostly follow the setup of Imhof et al~\cite{Imhof2005}. An exception 
will be when we discuss the extension of the model in Sec.~\ref{sec:IIPDE}. 
As such, we will consider a population of $N$ players with each player 
carrying out one of three pure strategies: AllC, AllD, or TFT. The respective 
payoffs resulting from an encounter of two players is characterized by the 
following payoff matrix:
\be
\bordermatrix{& AllC & AllD & TFT \cr
AllC & Rm & Sm & Rm  \cr
AllD & Tm & Pm & T+P(m-1) \cr
TFT & Rm-c & S+P(m-1)-c & Rm-c \cr},
\label{eqn:ipdmatrix}
\ee
where $m$ is the number of rounds played when two players meet. The parameters 
$T, R, P,$ and $S$ are the payoffs of the basic PD game (in which players meet 
only once): $T$ is the temptation to defect, i.e. the payoff a defector 
receives when playing a co-operator, $R$ is the reward for mutual cooperation, 
$P$ is the punishment for mutual defection, and $S$ is the sucker's payoff 
for co-operating with a defector. The so-called complexity cost, $c$, is 
imposed on the TFT strategy and represents the allocation of resources used to 
remember an opponent's last move~\cite{Binmore1992, Imhof2005}. For the dilemma 
to be present we require that the parameters satisfy $T>R>P>S$ and also that 
$R>(T+S)/2$, to prevent mutual alternate co-operation and defection being 
more profitable that of mutual cooperation~\cite{Axelrod1981}. Throughout this 
paper we use the specific parameter values $T=5$, $R=3$, $P=1$, $S=0.1$, 
and $m=10$~\cite{Imhof2005}. In the terminology of game theory, the iterated 
PD as defined by the above payoff matrix is a non-cooperative symmetric game.

In the following we will label the strategies AllC, AllD, and TFT by 
$i=1,2,3$ respectively. The number of players in the population using 
strategy $i$ will be denoted by $n_i$, and we require that $n_1+n_2+n_3=N$. 
The expected payoff, or fitness, of a player of type $i$ is then given 
by
\be\label{eqn:defpi}
\pi_i = \displaystyle \frac{\sum_j a_{ij}n_j-a_{ii}}{N-1},
\ee
where $a_{ij}$ are the elements of the payoff matrix, 
Eq.~(\ref{eqn:ipdmatrix}), e.g. $a_{11} = Rm$, $a_{12} = Sm$, etc. In using 
the definition (\ref{eqn:defpi}) we follow the choices of \cite{Nowak2006} 
and exclude interactions of one individual with itself. Definitions with 
self-interaction are possible, the differences do not affect the results to 
the order in inverse system size we will be working at.

The so-called reproductive fitness of an agent carrying pure strategy 
$i$, $f_i$, is defined as~\cite{Nowak2006}
\be
f_i = 1 - w + w\pi_i,
\label{eqn:payoff_fitness}
\ee
where $w$ is a selection strength that determines the impact that the game 
has on the agent's overall fitness. If $w=0$, then $f_i=1$ for all $i$, and 
one recovers the limit of neutral selection. For $w>0$ selection becomes 
increasingly frequency dependent. The average reproductive fitness in the 
population is then given by
\be
\phi = \displaystyle \sum_i \frac{n_i}{N}f_i,
\label{eqn:phi}
\ee
and the average payoff is $\pi=\sum_i (n_i/N)\pi_i$. In order to complete 
the model we need to specify the microscopic dynamics of the system, i.e. we 
need to define the rules by which the composition of the population changes 
over time. There are several such microscopic processes which have been 
studied in the literature, and we will define some of these in 
Sec.~\ref{sec:updaterules}. Before we do so, it will however be helpful to 
discuss the standard replicator-mutator dynamics commonly considered in 
the literature. Provided the microscopic dynamics are chosen appropriately, 
these equations are a suitable description in the deterministic limit, 
valid for infinite populations. It is however important to stress that 
the replicator-mutator dynamics are not the limiting deterministic dynamics 
for {\em all} microscopic processes, as 
pointed out in~\cite{Traulsen2005,Traulsen2006}, and as we will discuss 
in more detail below.

\subsection{Canonical replicator-mutator equation}
Within the standard replicator dynamics of evolutionary game theory the 
time evolution of the concentration of a strategy $i$ is given 
by~\cite{Taylor1978}
\be
\dot{x}_i = x_i(f_i^\infty - \phi^\infty),
\label{eqn:stndrep}
\ee
where $x_i$ denotes the concentration of strategy $i$ in the population in 
the deterministic limit: $x_i = \displaystyle \lim_{N\to\infty}n_i/N$. Similarly, 
we will write 
\be
f_i^\infty = \displaystyle \lim_{N\to\infty}f_i = 1 - w + w\sum_j a_{ij}x_j,
\ee
and
\be
\phi^\infty = \displaystyle \lim_{N\to\infty}\phi = \sum_j x_j f_j^\infty,
\label{eqn:phiinf}
\ee
where the superscripts indicate that Eq.~(\ref{eqn:stndrep})  are, for 
suitably chosen microscopic dynamics, valid only in the deterministic 
limit of infinite populations. The basic assumption underlying these 
dynamics is that individuals reproduce asexually in proportion to their 
reproductive fitness, and that offspring inherit the strategy of their 
parent.

If one introduces mutation, so that there is a finite chance that a player 
will produce an offspring which does not use the same strategy as their 
parent, the above dynamics needs to be modified, and the description is 
then in terms of so-called replicator-mutator equations~\cite{Nowak2001}. 
Focusing on the case of $M$ pure strategies we will assume that in a 
reproduction event a player produces an exact copy of itself with 
probability $1-(M-1)u$ and a mutant which plays one of the other $M-1$ 
strategies, each with probability $u$. The parameter $u$ is confined 
to the physically meaningful range $0 < u \leq 1/M$ for the case of $M$ 
pure strategies. For $u=1/M$ an offspring will be of any of the $M$ types 
with equal probability $1/M$.  It is convenient to introduce a mutation 
matrix
\begin{equation}
q_{ij} = \left\{ 
\begin{array}{l l}
  1-(M-1)u & \quad \mbox{if $i=j$}\\
  u & \quad \mbox{if $i\neq j$}\\ \end{array} \right..
\end{equation}
The replicator-mutator equation is then given by~\cite{Nowak2001}
\be
\dot{x}_i = \displaystyle \sum_j x_jf_j^\infty q_{ji} - x_i\phi^\infty.
\label{eqn:stndrepmut}
\ee
In the limit of zero mutation Eq.~(\ref{eqn:stndrepmut}) reduces to the 
standard replicator equation, Eq.~(\ref{eqn:stndrep}).

\subsection{Microscopic dynamics}
\label{sec:updaterules}
We will now define the different microscopic processes we will consider. We
will restrict ourselves to dynamics conserving the total number of players 
in the population. To specify a process it is then sufficient to define 
the `conversion' rates $T_{i\to j}$, corresponding to events in which 
a player of type $i$ is replaced by one of type $j$. For the general case 
with $M$ pure strategies $i,j=1,\dots,M$. We will limit the discussion to 
processes of the general form
\be
T_{i\rightarrow j} = \sum_k \frac{n_k}{N}\frac{n_i}{N}g_{ki}(\bm{f})q_{kj},
\label{eqn:transitions}
\ee
where $\bm{f} = (f_1, \dots, f_M)$. The form (\ref{eqn:transitions}) is found
by, at each time step, selecting two players, one for potential reproduction 
and one for potential removal, from the population. The player selected for
potential removal is assumed to be of type $i$, and each term in the sum 
corresponds to selecting a player of type $k$ for potential reproduction. A 
given combination $(i,k)$ thus occurs with probability $(n_i n_k)/N^2$. Here
we use sampling with replacement. Dynamics without replacement of an already 
chosen player are equally possible, leading to, for example, factors of 
$N(N-1)$ in the denominator instead of $N^2$. The differences amount to
effects of order $N^{-1}$, and do not affect results to the order in the 
inverse system size we will be working at. For a given pair of selected 
players reproduction and death actually only occur at a rate proportional 
to $g_{ki}(\bm{f})$, which here we assume to be a function of the reproductive 
fitnesses (implying a possible dependence on the average fitness $\phi$). 
The factor $q_{kj}$ accounts for potential mutation events. The four kinds 
of microscopic dynamics we will consider in the following differ in the 
details of the function $g$, which we will describe below. Choices in 
which $g_{ki}$ does not depend on $\bm{f}$ correspond to neutral selection.
 
Before we define the details of the different microscopic dynamics some 
general statements are appropriate. For simplicity, we will focus on 
the case of a game with $M=3$ pure strategies and in particular the 
iterated PD game with strategies AllC, AllD and  TFT  as introduced above; 
generalization to an arbitrary number of strategies $M$ is however 
straightforward. For any choice of $g_{ki}$, the state of the $N$-player 
population is defined by the number of individuals using the AllC and 
AllD strategies: $\bm{n}=(n_1,n_2)$, the number of TFT players is then given 
by $n_3 = N - n_1 - n_2$. Furthermore the reproductive fitnesses $\bm{f}$ 
and the average reproductive fitness, $\phi$, are fully determined by the 
state $\bn$ of the system (see Eqs.~(\ref{eqn:defpi})-(\ref{eqn:phi})). It 
follows that the transition rates $T_{i\to j}$ can be written as functions 
of $\bm{n}$, and the microscopic stochastic process is described by the 
following master equation for the probability, $P(\bm{n},t)$, of the 
system being in state $\bm{n}$:
\begin{eqnarray}
\label{eqn:mastertwo}
\frac{dP(\bm{n},t)}{dt} &=& (\shifta-1)T_{1\rightarrow 3}(\bm{n})P(\bm{n},t) \\
&+& (\shiftb-1)T_{2\rightarrow 3}(\bm{n})P(\bm{n},t) \nonumber \\
&+& (\shifta\shiftb^{-1}-1)T_{1\rightarrow 2}(\bm{n})P(\bm{n},t) \nonumber \\
&+& (\shiftb\shifta^{-1}-1)T_{2\rightarrow 1}(\bm{n})P(\bm{n},t) \nonumber \\
&+& (\shifta^{-1}-1)T_{3\rightarrow 1}(\bm{n})P(\bm{n},t) \nonumber \\
&+& (\shiftb^{-1}-1)T_{3\rightarrow 2}(\bm{n})P(\bm{n},t). \nonumber
\end{eqnarray}
Here we have introduced shift operators $\shifti$, where $i = 1,2$, acting 
on functions of the state of the system, $\psi(n_1,n_2)$, as follows:
\BE
\shifta \psi(n_1, n_2) &=& \psi(n_1 + 1, n_2), \nonumber \\
\shifta^{-1} \psi(n_1, n_2) &=&\psi(n_1 - 1, n_2).
\EE
Similar definitions apply for $\shiftb$ and $\shiftb^{-1}$. Multiplying 
both sides of Eq.~(\ref{eqn:mastertwo}) by $\bm{n}$ and summing over all 
possible states and using a decoupling approximation, valid for 
$N\to\infty$, we can then write the rate of change of the concentration 
of strategy $i$ as
\be
\dot{x}_i = \sum_{k\neq i} \left[ T_{k\rightarrow i}(\bm{x}) - 
T_{i\rightarrow k}(\bm{x}) \right],
\label{eqn:genrepmut}
\ee
after a re-scaling of time by a factor $N$. Clearly this equation is not 
restricted to the case $M=3$, and holds when an arbitrary number of 
strategies are present. Transition rates in Eq.~(\ref{eqn:genrepmut}) are 
found from Eq.~(\ref{eqn:transitions}) using the substitution $n_i/N\to x_i$ 
and $f_i \to f_i^{\infty}$, see Appendix A for further details. Substituting 
these limits into Eq.~(\ref{eqn:genrepmut}) one finds that the deterministic 
evolution of the concentrations of strategies is given by
\be
\dot{x}_i = \sum_{k \neq i} \sum_{j} x_j \left[ x_k g_{jk} q_{ji} -
x_i g_{ji} q_{jk} \right].
\label{eqn:genrepmut2}
\ee
For different update rules this equation differs only in the specific form 
of $g$ used.

We will now proceed to give the specific form of the function 
$g_{ki}(\bm{f})$ for a set of different update rules which have previously 
been proposed: the Moran process, a linear Moran process, a local process 
and the Fermi process~\cite{Claussen2007,Traulsen2006}.

\subsubsection{The Moran process}
In the Moran process~\cite{Moran1958}, once a player of type $k$ has been 
chosen for potential reproduction and a player of type $i$ for potential 
removal, the reproduction event occurs at a rate proportional to $f_k/\phi$, 
specifically we will choose
\be\label{eqn:gmoran}
g^M_{ki}(\bm{f}) = \frac{f_k}{2\phi}.
\ee
The arbitrary pre-factor of $1/2$, equivalent to choosing a time scale, has 
been introduced to allow better comparison with other update 
rules~\cite{Claussen2007}.  By substituting Eq.~(\ref{eqn:gmoran}) into 
Eq.~(\ref{eqn:genrepmut2}) and using Eq.~(\ref{eqn:phiinf}), $\sum_k x_k = 1$,
and $\sum_k q_{jk} = 1$, one finds specifically for the Moran process that
\be
\dot{x}_i = \displaystyle \frac{\sum_j x_jf_j^\infty q_{ji} - 
x_i\phi^\infty}{2\phi^\infty}.
\label{eqn:morandet}
\ee
It is important to stress that the average reproductive fitness $\phi^\infty$ 
is a function of the concentration vector $\bm{x}$, and so $\phi^\infty$ is 
a time-dependent quantity.  While Eq.~(\ref{eqn:morandet}) is similar to the 
standard replicator-mutator dynamics, the pre-factor $(2\phi^\infty)^{-1}$ 
corresponds to a {\em dynamic} re-scaling of time, and so may affect the 
transient dynamics. The location of fixed points and their local stability 
are however not affected, as a straightforward calculation shows. 

\subsubsection{The linear Moran process}
The linear Moran process is defined by the following choice~\cite{Claussen2007} 
\be\label{eqn:glm}
g_{ki}^{LM}(\bm{f}) = \frac{1}{2}\left(1+\lambda(f_k-\phi)\right),
\ee
where $\lambda>0$ is a constant parameter, such that it is always the case 
that  $T_{i\to j} \geq 0$. Notice also that one has 
$g_{ki}^{LM}(\bm{f})=(1+\lambda w(\pi_k-\pi))/2$. A common choice, which 
we will adopt in the following, is $\lambda=1/\Delta f_{max}$, where 
$\Delta f_{max}$ is the maximum possible difference between $f_i$ and 
$\phi$~\cite{Claussen2007}, i.e. 
$\Delta f_{max}=\max_{k,\bn} |f_k(\bn)-\phi(\bn)|$. In the absence of mutation 
($u=0$) the deterministic limit results in the following dynamics
\be\label{eqn:replm}
\dot x_i=\frac{x_i(f_i^\infty-\phi^\infty)}{2 \Delta f_{max}}.
\ee
Therefore, up to a re-scaling of time by the constant factor 
$(2\Delta f_{max})^{-1}$, the linear Moran process without mutation is 
described by the standard replicator dynamics in the limit of infinite 
population size. However for $u \neq 0$ one does not recover the standard 
replicator-mutator equations, Eq.~(\ref{eqn:stndrepmut}), from the linear 
Moran process. 

In both Eq.~(\ref{eqn:replm}) and (for $u \neq 0$) from the result of 
substituting Eq.~(\ref{eqn:glm}) into Eq.~(\ref{eqn:genrepmut2}), the 
reproductive fitness only enters in differences of the form $f_k-\phi$ or 
$f_k-f_i$ and is normalized by $\Delta f_{max}$. Since both, fitness 
differences and $\Delta f_{max}$, scale linearly in $w$, the deterministic 
dynamics is independent of the selection strength $w$ for the linear Moran 
process. Finally, the linear Moran process can be obtained from Moran 
process Eq.~(\ref{eqn:gmoran}) in the weak selection limit, $w\ll 1$. Using 
Eq.~(\ref{eqn:payoff_fitness}) one has
\begin{eqnarray}
g_{ki}^M(\bm{f}) &=& \frac{1-w+w\pi_k}{2(1-w+w\pi)} \nonumber \\
&=& \frac{1}{2}\left(1+w(\pi_k-\pi)\right)+{\cal O}(w^2),
\label{eqn:mlin}
\end{eqnarray}
so that to linear order one recovers Eqs.~(\ref{eqn:glm}) with the choice 
$\lambda=1$.

\subsubsection{The local process}
The so-called local process was first proposed by Traulsen et 
al~\cite{Traulsen2005} and is based on a pairwise comparison of one agent's 
fitness with another in order to determine whether or not reproduction occurs. 
This process has the advantage that no knowledge or computation of the average 
fitness of the population is required to carry out a microscopic step. The 
local process is defined by
\be\label{eqn:llin}
g_{ki}^L(\bm{f}) = 
\frac{1}{2}\left(1 + \frac{f_k-f_i}{\Delta f_{max}}\right),
\ee
where $\Delta f_{max}$ is again required for normalization and fixed at the 
beginning, and then remains unchanged as the dynamics proceeds. As opposed 
to the case of the linear Moran process, $\Delta f_{max}$ is now the maximum 
possible absolute difference between any two fitnesses $f_i$ and $f_k$: 
$\Delta f_{max}=\max_{i,k,\bn} |f_i(\bn)-f_k(\bn)|$. As with the linear Moran 
process the local process does, up to a constant factor which can be absorbed 
in the definition of time, reproduce the standard replicator equation 
(\ref{eqn:stndrep}) in the deterministic limit if mutation is 
absent~\cite{Traulsen2005,Traulsen2006}. At finite mutation rates one does 
not however recover the replicator-mutator equation, 
Eq.~(\ref{eqn:stndrepmut})~\cite{Traulsen2006}. 

\subsubsection{The Fermi process}
Finally, the so-called Fermi process is an alternative pairwise comparison 
process which uses the Fermi-Dirac distribution instead of the linear 
functional dependence on fitness differences as in the local process. It is
defined by~\cite{Traulsen2006,Claussen2008}
\be
g_{ki}^F(\bm{f}) = \frac{1}{2}\left[1 + \tanh
\left( \frac{1}{2}\left(f_k - f_i\right) \right) \right].
\label{eqn:gfermi}
\ee
Unlike the other update rules, the Fermi process does not reproduce either 
the standard replicator or replicator-mutator equations in the deterministic 
limit. From Eq.~(\ref{eqn:gfermi}) one has
\be
g_{ki}^F(\bm{f})=\frac{1}{2}
\left(1+\frac{w}{2}(\pi_k-\pi_i)\right)+{\cal O}(w^2),
\ee
in the limit of weak selection. This is of the same functional form as 
Eq.~(\ref{eqn:llin}).

\section{Results of the deterministic analysis}
\label{sec:IPDdeterministic}
As an initial step towards characterizing the outcome of the iterated PD, we 
compute the fixed-point structure in the deterministic limit of the four 
different dynamics defined above, as function of the complexity cost $c$ 
and the mutation rate $u$. We fix the selection strength to $w=1$ throughout.
 

\begin{figure}
\vspace{2em}
\centering
\includegraphics[scale=0.4]{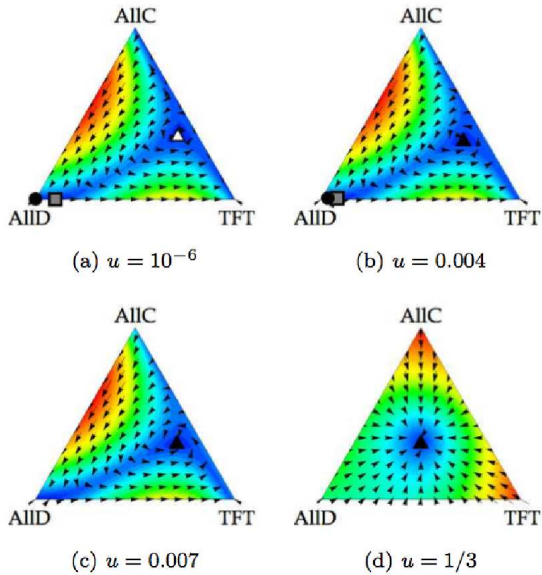}
\caption{(Color online) Fixed-point structure and flow fields of the 
standard replicator-mutator equations for the iterated PD at $c=0.8$ and 
$w=1$. Black symbols are stable fixed points, white symbols are unstable 
and gray symbols denote saddle points. The AllD fixed point and saddle point 
can be seen in the bottom left-hand corner of the simplices. The mixed fixed 
point (triangle) changes stability at $u_c^{(1)} \approx 0.0016$. The AllD 
fixed point (circle) and the saddle point (square) annihilate at 
$u_c^{(2)} \approx 0.005$. Arrows indicate the direction of the deterministic 
flow in the strategy simplex. The color map shows the Euclidean speed of 
the trajectories, $||\dot{\bm{x}}||$. These images were produced using a 
modified version of the Dynamo package \cite{Dynamo2007}.}
\label{fig:M_simplexes}
\end{figure}


\subsection{General fixed point structure}
The qualitative picture one finds is similar for any of the four dynamics; 
two different threshold values of the mutation rate can be identified, we 
will refer to these as $u_c^{(1)}$ and $u_c^{(2)}$.  For $0<u<u_c^{(1)}$, 
one typically finds three fixed points: a locally stable attractor near 
AllD, a saddle point also near AllD, and an unstable fixed point, located 
close to the AllD/TFT edge of the strategy simplex, see 
Fig.~\ref{fig:M_simplexes}(a). Following~\cite{Imhof2005} we will refer to 
this latter fixed point as the `mixed fixed point'. At $u= u_c^{(1)}$ the 
mixed fixed point becomes stable, as shown in panel (b) of 
Fig.~\ref{fig:M_simplexes}. At $u = u_c^{(2)}$, the two fixed points near 
AllD annihilate, leaving the mixed fixed point as the only attractor for 
$u > u_c^{(2)}$, see Fig.~\ref{fig:M_simplexes}(c). At $u = 1/3$ the mixed 
fixed point is at or near the center of  the simplex, see 
Fig.~\ref{fig:M_simplexes}(d). At this maximal physical meaningful value 
of $u$ an individual of any type produces an offspring of any of the 
three different strategies with equal probability. While this qualitative 
picture is the same for all four dynamics considered here, the numerical 
values of $u_c^{(1)}$ and $u_c^{(2)}$ will in general be different for 
the different dynamics, and they may also depend on the choice of the 
complexity cost, $c$.  The overall picture is consistent with the results 
of~\cite{Imhof2005}, where the standard replicator dynamics were studied and 
where similar qualitative behavior was found. Our analysis thus demonstrates 
that the findings of~\cite{Imhof2005} generalize to a broader class of 
dynamics. The only difference between our findings compared to those of 
earlier analyses, lies in the saddle point described above, which was not 
reported in~\cite{Imhof2005}, presumably because it is not an attractor 
of the dynamics. Although for completeness we have given a general account 
of the fixed point structure, the mixed fixed point will be the focus of 
our analysis in the following sections, as it is this fixed point which 
gives rise to coherent oscillations induced by demographic noise.

\subsection{Limit of small mutation rates}
Further analytical progress is possible in the limit of small mutation 
rates, $u\ll 1$. For all four cases considered here the deterministic 
dynamics, derived from Eqs.~(\ref{eqn:genrepmut}), are of the form 
\be
\label{eqn:sep}
\dot{\bm{x}} = \bh_{(0)}(\bm{x})+u\bh_{(1)}(\bm{x}),
\ee
with the mutation rate entering linearly in the resulting differential 
equations. We will now make the following ansatz for a fixed point $\bm{x}^*$:
\be
\bm{x}^* = \bm{x}_{(0)}^* + u\bm{x}_{(1)}^* ,
\label{eqn:linearansatz}
\ee
in the limit of small mutation rates $u$. Here $\bm{x}_{(0)}^*$ is a fixed 
point of Eq.~(\ref{eqn:genrepmut}) at $u=0$ and $\bm{x}_{(1)}^*$ captures the 
effect of non-zero mutation rates at next-to-leading order. Inserting 
Eq.~(\ref{eqn:linearansatz}) into the fixed-point condition 
\be\label{eqn:fpcond}
\bh_{(0)}(\bm{x}^*) + u\bh_{(1)}(\bm{x}^*)=0,
\ee
and collecting terms in linear in $u$, one then finds
\be
\bm{x}_{(1)}^*=-J^{-1}\bm{h}_{(1)} (\bm{x}_{(0)}^*),
\label{eqn:fplinmutdev}
\ee
where $J$ is the Jacobian of $\bh_{(0)}$ evaluated at $\bm{x}^*=\bm{x}_{(0)}^*$.
Since $\bm{x}^*_{(0)}$ can be found in closed form from 
Eqs.~(\ref{eqn:genrepmut}) with $u=0$, substituting, 
Eq.~(\ref{eqn:fplinmutdev}) into Eq.~(\ref{eqn:linearansatz}) gives an 
analytical prediction of the location of the fixed point at small $u$. 

\begin{figure}
\vspace{2em}
\centering
\includegraphics[scale=0.3]{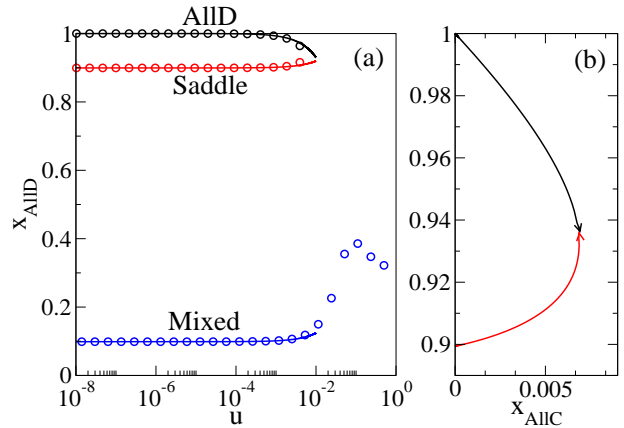}
\caption{(Color online) Panel (a) shows the $x_{AllD}$ components of the 
three fixed points of the Moran dynamics at $c=0.8$ with changing mutation 
rate $u$. Circles are the results from numerically solving the fixed-point 
equations (\ref{eqn:fpcond}), lines are from
Eq.~(\ref{eqn:linearansatz}). Panel (b) shows the paths of the AllD fixed 
point and saddle point in phase space as $u$ is varied. The fixed points 
meet and annihilate at $u\approx 0.005$, for $u$ larger than this value 
neither fixed point is present.}
\label{fig:M_fptrack}
\end{figure}

In Fig.~\ref{fig:M_fptrack} we compare the outcome of the above linear 
expansion with results from a direct numerical evaluation of the fixed points 
of Eq.~(\ref{eqn:genrepmut}) obtained using a Newton-Raphson procedure. The 
expansion is seen to be a good approximation for the location of the fixed 
point for values of $u$ up to $u\approx 0.01$. From the figure we see that 
the AllD and saddle fixed points annihilate at $u_c^{(2)}\approx 0.005$ for 
this value of $c$. This annihilation is consistent with the disappearance 
of the AllD fixed point at large $u$ reported by Imhof et al~\cite{Imhof2005}.


\subsection{Mixed fixed point and phase diagram}

\begin{figure}
\vspace{2em}
\centering
\includegraphics[scale=0.3]{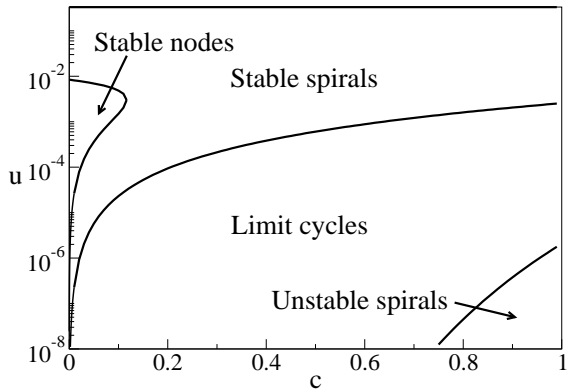}
\caption{A phase diagram showing the regions in the $(c,u)$ plane for which 
the mixed fixed point of Eq.~(\ref{eqn:genrepmut}) for the Moran process is 
a stable node, stable spiral, unstable spiral or is orbited by a limit cycle. 
The solid line at $u=1/3$ indicates that at this value the mixed fixed point 
becomes a stable node situated at $\bm{x} = (1/3,1/3)$. Note that due to 
the procedure used to check for the presence of limit cycles, the border 
between the limit cycle and unstable spiral regions is only approximate.}
\label{fig:M_heatmap}
\end{figure}

For suitable choices of the model parameters $c$ and $u$, the mixed fixed 
point can be a stable attractor with complex eigenvalues of the 
corresponding Jacobian. One can thus expect coherent stochastic oscillations 
to arise in finite populations at those model parameters. We therefore 
focus our attention on the mixed fixed point, and identify the regions in 
the $(c,u)$-plane where such complex eigenvalues are found. More generally 
we will determine the nature of the mixed fixed point as a function of $u$ 
and $c$. The result of numerically solving for fixed points of the 
deterministic dynamics corresponding to the Moran process is shown in 
Fig.~\ref{fig:M_heatmap}. We will denote fixed points with purely real 
eigenvalues as `nodes' and those with complex eigenvalues as `spirals'. At 
low values of $c$ we observe a re-entry phenomenon, where the mixed fixed 
point goes from a stable spiral to a stable node and back to a stable spiral 
as $u$ is decreased. 

Numerically we also observe a region where the dynamics converges onto a 
limit cycle. We are at this point unable to provide a proof for the 
existence of limit cycles or to analytically determine the position of the 
border between the limit cycle and unstable spiral regions. We therefore 
determine the presence of limit cycles by numerically integrating the 
deterministic dynamics using an Euler forward method, starting from the 
center of the simplex, allowing for a period of equilibration and then 
applying a suitable threshold criterion to detect closed trajectories. The 
unstable spiral region is identified as the region where we do not find 
limit cycles numerically. In situations where there is more than one 
attractor (e.g. a limit cycle and a stable fixed point near AllD) initial 
conditions will determine the stationary state of the dynamics. At $u=1/3$ 
the mixed fixed point is located in the center of the strategy simplex, and 
becomes a stable node.


\begin{figure}
\vspace{2em}
\centering
\includegraphics[scale=0.3]{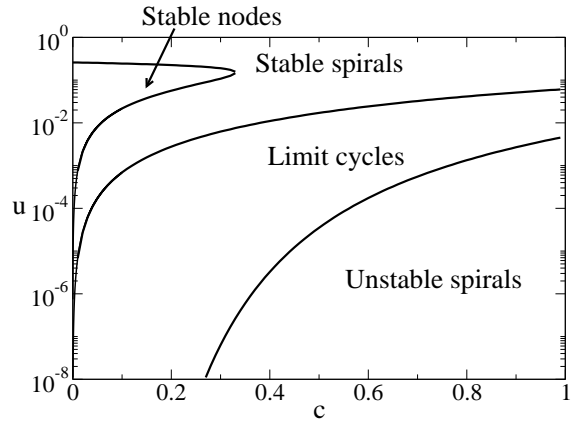}
\caption{Phase diagram for the Fermi process. We again see the same 
qualitative structure as for the other update rules, with a difference only 
in the positions of the boundaries.}
\label{fig:F_heatmap}
\end{figure}

All other update rules studied in this paper have the same qualitative 
features as the Moran process, and hence their phase diagrams are 
structurally similar to that shown in Fig.~\ref{fig:M_heatmap}, except that 
the mixed fixed point does not become a stable node at $\bm{x}=(1/3,1/3)$ 
at $u = 1/3$ for rules that use pairwise comparison. Instead, the fixed 
point forms a stable spiral close to the center of the simplex. Although 
qualitative features of the phase diagrams are the same for all four update 
rules, the quantitative positions of the borders in the $(c,u)$ plane may 
differ for each update rule. For example, Fig.~\ref{fig:F_heatmap} shows the 
stability map for the Fermi process. Here the re-entry region persists for 
larger values of $c$ and the region in which the mixed fixed point is unstable 
is also much larger.  


\section{Stochastic effects and system-size expansion}
\label{sec:IPDstoc}
\begin{figure}
\vspace{2em}
\centering
\includegraphics[scale=0.3]{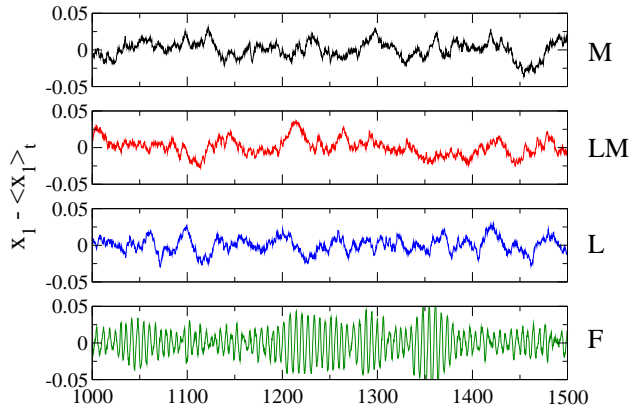}
\caption{(Color online) The results for the concentration of the 
AllC strategy from one run of a Gillespie 
simulation~\cite{Gillespie1976,Gillespie1977} for each of the four update rules at $N=10000$, $c=0.8$, 
and $u=0.05$. The time averaged concentration of each run has been subtracted 
from the data to give the deviation from the deterministic fixed point.}
\label{fig:gilcomps}
\end{figure}

Until now we have focused on the dynamics of infinite populations. In this
section we investigate effects arising in finite populations, especially 
stochastic oscillations arising via a coherent amplification of intrinsic 
fluctuations. Such oscillations have been found in a variety of systems as 
described in the introduction. These quasi-cycles typically arise in 
regions of the phase diagram where the deterministic dynamics approach a 
fixed point, and so the range of parameters in which systems of finite 
populations display oscillations is generally wider than the region in which 
the deterministic system allows for periodic solutions. Fig.~\ref{fig:gilcomps}
indeed confirms that this is also the case for the evolutionary dynamics 
of the iterated PD. In the figure we choose model parameters such that none of the four deterministic dynamics approach periodic attractors, 
but instead have stable fixed points with complex eigenvalues (stable 
spirals). As seen in the figure the dynamics in finite populations still 
generate oscillatory behavior, induced by intrinsic fluctuations. This 
oscillatory behavior is similar to that reported in~\cite{Imhof2005}. The 
four panels demonstrate that the quality and frequency of these stochastic 
oscillations can vary over a wide range depending on the details of the 
microscopic dynamics, and so we will now go on to characterize their
properties in more detail in order to obtain a more comprehensive picture of 
this phenomenon.


The analytical approach we will use to characterize these fluctuations 
is based on an expansion of the master equation in the inverse system 
size~\cite{Kampen2007}. This method is now standard in the analysis of 
interacting-agent systems, and we will therefore not present the full 
details of the calculation, but instead restrict ourselves to giving 
a few of the intermediate steps and the final results. The starting point 
of the system-size expansion is an ansatz of the type 
\be
\label{eqn:decomp}
\frac{n_i}{N} = x_i(t) + \frac{1}{\sqrt{N}}\xi_i(t),
\ee
amounting to a separation of deterministic and stochastic contributions to 
the number, $n_i$, of individuals of type $i$ in the population. The first 
term on the right, $x_i(t)$, is the deterministic trajectory, and $\xi_i(t)$ 
captures fluctuations about this trajectory; the magnitude of these 
fluctuations is expected to be of order $N^{-1/2}$, as reflected by the above 
ansatz. One proceeds by inserting this ansatz into the master equation 
(\ref{eqn:mastertwo}) and focuses on the probability distribution of 
$\bm{\xi}$, rather than that of $\bn$, so that one sets 
$P(\bm{n},t) = \Pi(\bm{\xi},t)$. Expanding the resulting master equation for 
$\Pi(\bm{\xi},t)$ in powers of $N^{-1/2}$ one recovers, at leading order, the 
generalized replicator-mutator equation, Eq.~(\ref{eqn:genrepmut}). At 
next-to-leading order in $N^{-1/2}$ a Fokker-Planck equation of the 
form
\be
\pd{\Pi}{t}=-\sum_i\pd{}{\xi_i}(C_i\Pi)+
\frac{1}{2}\sum_{i,j}B_{ij}\pd{^2\Pi}{\xi_i\partial{\xi_j}},
\label{eqn:fokkerplanck}
\ee 
is found~\cite{Kampen2007}, where $C_i = \sum_k J_{ik}\xi_k$. Here $J$ is 
the Jacobian of Eq.~(\ref{eqn:genrepmut}) and $B$ is a symmetric, 
$2\times 2$ matrix, whose precise form will depend on the exact nature of 
the microscopic dynamics, but whose general form is given in 
Appendix~\ref{app:vankampen}. The Fokker-Planck 
equation (\ref{eqn:fokkerplanck}) is equivalent to the linear Langevin 
equation~\cite{Gardiner1985}
\be
\dot{\bm{\xi}} = J\bm{\xi} + \bm{\eta},
\label{eqn:langevin}
\ee
where $\bm{\eta}$ is Gaussian white noise with correlations
\be
\left<\eta_i(t)\eta_j(t^{\prime})\right> = B_{ij}\delta(t-t^{\prime}).
\ee
In contrast to the Langevin equations derived using the Kramers-Moyal 
expansion \cite{Traulsen2006}, Eq.~(\ref{eqn:langevin}) contains 
additive, rather than multiplicative noise. In the application we are 
considering here, we are interested in fluctuations about the stationary 
state and so the matrices $J$ and $B$ are evaluated at the fixed point of 
the deterministic dynamics.

Given the linearity of Eq.~(\ref{eqn:langevin}), it is straightforward to 
compute the power spectra of the fluctuations $\bm{\xi}$ about the 
deterministic fixed point. Following the steps of~\cite{McKane2007}, one 
obtains
\be
P_i(\omega) = \left<|\widetilde{\xi}_i(\omega)|^2\right> = 
\sum_j\sum_k \Phi_{ij}^{-1} B_{jk} (\Phi^\dagger)^{-1}_{ki},
\label{eqn:ps}
\ee
where $\Phi = i\omega\mathbb{I}-J$ and $\mathbb{I}$ is the $2\times 2$ 
identity matrix.

\begin{figure}
\vspace{2em}
\centering
\includegraphics[scale=0.3]{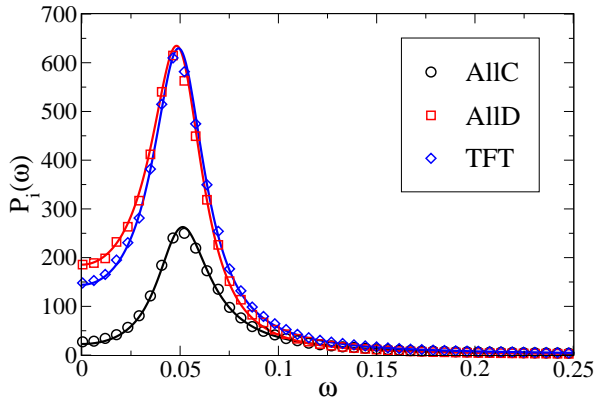}
\caption{(Color online) The power spectra for oscillations in the 
concentrations of the three strategies with Moran updating for $c=0.8$ and
$u=0.01$. Symbols are the results of a Gillespie simulation with $N=12000$ 
and approximately $10^4$ runs. Solid lines are theoretical predictions 
obtained from Eq.~(\ref{eqn:ps}). Simulation results show excellent 
agreement with the theory.}
\label{fig:M_ps}
\end{figure}

In Fig.~\ref{fig:M_ps} we give an example of the power spectra of fluctuations 
about the deterministic fixed point obtained for the Moran update rule at 
$c=0.8$ ($w=1$), $u=0.01$, and $N=12000$. Theoretical predictions from the 
van Kampen expansion and numerical simulations are in near perfect agreement. 
The spectrum shows a pronounced peak at a frequency of approximately 
$\omega=0.05$, indicating the existence of amplified oscillations with that 
characteristic frequency. The amplitude of these oscillations is proportional to $N^{-1/2}$, see Eq.~(\ref{eqn:decomp}), the proportionality constant is determined by the area under the power spectrum. Depending on the choice of parameter values one can then expect the amplitude of the quasi-cycle will be of order one up to system sizes of $10^4$ or so, i.e. comparable to the species concentrations 
at the deterministic fixed point. Even for very large populations the 
oscillations can therefore be significant. If the trajectory of the system is monitored over a time scale much smaller than the oscillation period, then this may lead to intervals in time in which the concentration of AllC is found to be consistently higher than that of TFT or AllD, that is, to intermediate periods where co-operation dominates the population. Such effects may for example be relevant when evolutionary time scales are much longer than time windows over which measurements can be made.


\begin{figure}
\vspace{2em}
\centering
\includegraphics[scale=0.3]{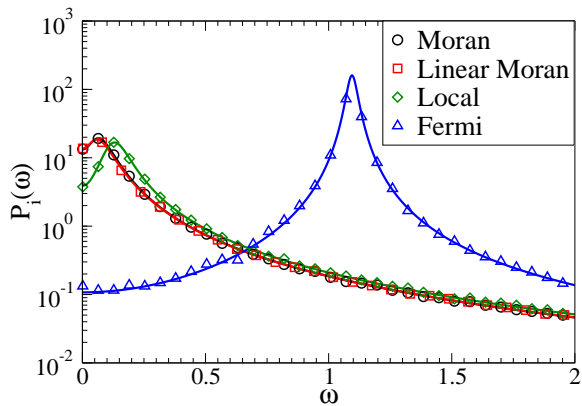}
\caption{(Color online) A comparison of the power spectra for oscillations 
in the AllC strategy concentration at $c=0.8$ and $u=0.05$. $N$ ranges from 
$10^4$ to $10^6$ and the number of runs for each simulation is of order 
$10^4$.}
\label{fig:pscomp}
\end{figure}

Having shown that the analytical approach captures the properties of 
quasi-cycles accurately, we can now compare the magnitude of the stochastic 
oscillations for the different update processes at the same values of $c$ 
and $u$. The power spectra of the fluctuations in the AllC concentration 
are shown in Fig.~\ref{fig:pscomp} for the four update rules at one fixed 
mutation rate and for a specific choice of the complexity cost. Results 
indicate that the Fermi process produces demographic oscillations of a 
higher frequency than the other update rules, in line with the time series 
shown in Fig.~\ref{fig:gilcomps}. Even though the power spectra for the 
Moran and linear Moran update rules are seemingly indistinguishable in 
Fig.~\ref{fig:pscomp}, they are not analytically equivalent. 


The magnitude of the peak in the power spectra is a good proxy for the 
amplitude of the stochastic quasi-cycles, and the height of the peak is in 
turn largely determined by the inverse of the real part of the relevant 
eigenvalue of the deterministic dynamics at the fixed point. In the 
deterministic system, perturbations about the fixed point decay with a time 
constant proportional to the inverse of this real part, and it is intuitively 
easy to see that the magnitude of stochastic oscillations diverges as the 
real part of the largest eigenvalue tends to zero. More specifically, as 
shown in~\cite{Boland2008}, the magnitude of the peak in the spectra diverges 
as the system approaches a Hopf bifurcation, where the stable spiral becomes 
an unstable one. The resulting delta-function peak in the power spectrum 
indicates that a limit cycle is born in the unstable phase. This can also 
be seen from Eq.~(\ref{eqn:ps}) and the definition of the matrix $\Phi$. 
At the Hopf bifurcation the relevant eigenvalue of $J$ is purely imaginary, 
and when $\omega$ becomes equal to the imaginary part of this eigenvalue, 
the matrix $\Phi$ becomes singular, such that the expression on the 
right-hand side of Eq.~(\ref{eqn:ps}), involving the inverse of $\Phi$, 
diverges.

\begin{figure}
\vspace{2em}
\centering
\includegraphics[scale=0.3]{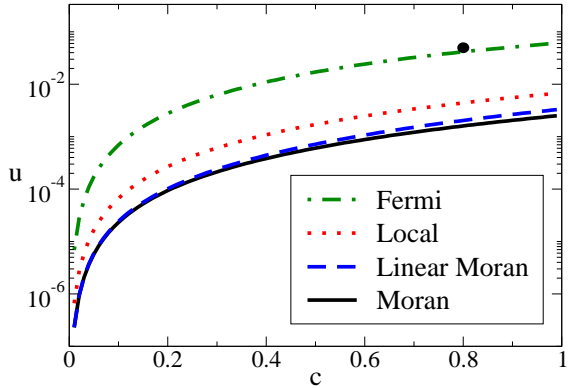}
\caption{(Color online) The borders between the stable spiral region 
(above the respective lines) and limit cycle region (below) for the 
different update processes. The black dot indicates the point 
$(c,u)=(0.8,0.05)$ used in Fig. 7.}
\label{fig:sslcbordercomp}
\end{figure}

In order to compare the relative magnitude of stochastic oscillations in 
the four different dynamics at fixed values of $u$ and $c$, it is therefore 
useful to determine how far or near to the instability the pair $(u,c)$ 
places the respective dynamics. In Fig.~\ref{fig:sslcbordercomp} we plot 
the instability lines indicating the occurrence of a Hopf bifurcation in 
the $(u,c)$ plane for the four different types of dynamics. For any fixed 
$c$ one finds that $u_{c,F}^{(1)}>u_{c,L}^{(1)}>u_{c,LM}^{(1)}>u_{c,M}^{(1)}$ 
and that accordingly for any $u$ sufficiently large to place all four 
dynamics in the stable regime, the Fermi process is much closer to the 
limit-cycle regime than the other types of dynamics, and would therefore 
be expected to have a larger peak in the power spectra. As discussed above 
we furthermore find that the Fermi process, with its alternative form of 
the pairwise comparison process, produces demographic oscillations of a 
higher frequency than the other update rules, see Fig.~\ref{fig:pscomp}.


\section{Iterated prisoner's dilemma with errors}
\label{sec:IIPDE}

In this section we study an extended version of the iterated PD game, 
allowing for a fourth pure strategy, win-stay lose-shift (WSLS). It is 
appropriate to include this strategy in the discussion when so-called  
`trembling-hand' errors are considered~\cite{Imhof2007}. Trembling-hand 
errors introduce the possibility of a player making a mistake after they 
have decided what to play, that is, a player co-operating when they meant 
to defect, or defecting when the intention was to co-operate. We will 
assume that the two players make errors of this type independently with a 
small probability $\epsilon>0$ in any given round. TFT's disadvantage is 
then that it can become locked into a cycle of alternate co-operation and 
defection with another TFT player after a mistake occurs. In such games, 
TFT can be outperformed by WSLS~\cite{Nowak2006}. WSLS co-operates initially 
and then keeps using its strategy (co-operation or defection) whenever 
it receives payoff $T$ or $R$ and switches its strategy (from co-operation 
to defection or vice versa) if it receives $P$ or $S$. WSLS does not become 
locked in such cycles when playing against TFT or WSLS. We include the 
WSLS strategy in our game, extending the dynamics to three degrees of 
freedom. 

In the presence of trembling-hand errors the outcome of a PD game between 
two fixed players and iterated for a finite number of rounds will generally 
be stochastic and depend on the timing at which errors occur in the 
interaction sequence. In order to simplify matters we will therefore 
follow~\cite{Imhof2007} and restrict the discussion to cases in which an 
interaction between two players consists of an infinite number of 
iterations of the PD game. It is then appropriate to use the expected 
payoffs per round, i.e. for two fixed players, say of types 
$i,j\in\{\mbox{AllC, AllD, TFT, WSLS}\}$, one formally considers an infinite 
sequence of PD interactions, and uses the mean payoff per round to define 
the payoff matrix elements $a_{ij}$. The payoff matrix can then be 
worked out for small error rates, and is given in~\cite{Imhof2007} and 
reproduced in Appendix B for convenience. The complexity cost, $c$, is no 
longer a relevant parameter now that the number of rounds is infinite.

\begin{figure}
\vspace{2em}
\centering
\includegraphics[scale=0.3]{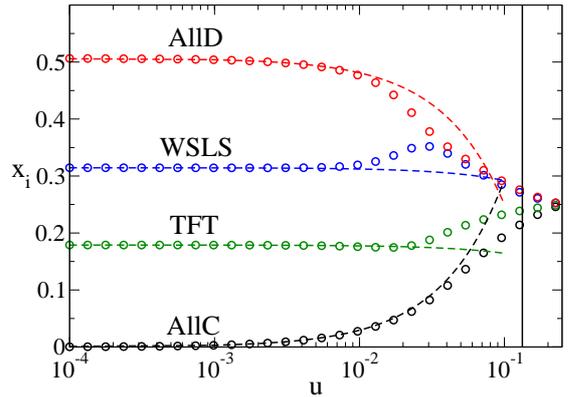}
\caption{(Color online) The AllC, AllD, TFT and WSLS components of the mixed 
fixed point for the infinitely iterated PD for changing $u$ at 
$\epsilon = 10^{-4}$. Microscopic dynamics are of the Moran type. Dashed
lines are analytical results from a linear expansion in $u$, circles are 
from a numerical evaluation of the fixed point. The vertical solid black 
line indicates the location of $u_c^{(1)}$, i.e. the value of $u$ where the 
mixed fixed point changes stability.}
\label{fig:IIPDE_M_fptrack}
\end{figure}

Previous work on this game has shown that in the limits of zero mutation
and weak selection the population can either fix on WSLS or AllD depending 
on the values used in the payoff matrix \cite{Imhof2007}. We continue to 
use the parameter values given in Sec.~\ref{sec:IPDmodel} and explore how 
the dynamics of the game depend on mutation and error rates and identify 
and classify demographic oscillations. Analyzing the four update rules 
given in Sec.~\ref{sec:updaterules} we again find a mixed fixed point on 
which we focus our analysis --- since demographic oscillations may occur 
about this fixed point when it is a stable spiral. We use 
Eq.~(\ref{eqn:genrepmut}) with four strategies to track the location and 
stability properties of the mixed fixed point as $u$ is varied. The path 
of the mixed fixed point at constant $\epsilon$ and changing $u$ for the 
Moran process, is shown in Fig.~\ref{fig:IIPDE_M_fptrack}. The dashed 
lines are the result of a similar perturbative expansion to that carried 
out in Sec.~\ref{sec:IPDdeterministic}, where again we see good agreement 
with numerical results for changes in $u$ up to $u\approx 0.01$. 


\begin{figure}
\vspace{2em}
\centering
\includegraphics[scale=0.3]{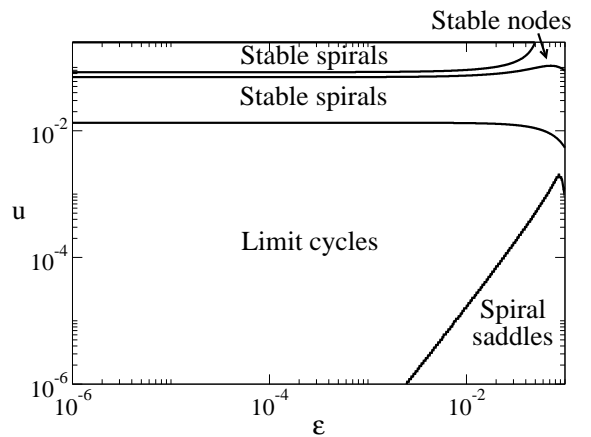}
\caption{The stability of the mixed fixed point for Moran updating in the 
infinitely iterated PD game in the $(\epsilon,u)$ plane. Note that the
border between the limit cycle and spiral saddle regions is approximate.}
\label{fig:IIPDE_M_heatmap}
\end{figure}


Similar to our analysis of the three-strategy game, we can determine the 
stability of the mixed fixed point as a function of the model parameters 
$\epsilon$ and $u$. The classification of the nature of the fixed points is 
more involved for the four-strategy game, however, as we are analyzing a 
three-dimensional dynamical system. Stable spirals are now fixed points 
with one pair of complex-conjugate eigenvalues with a negative real part 
and an additional real-valued negative eigenvalue. If the sign of the real 
part of the pair of complex-conjugate eigenvalues is opposite to that of 
the real-valued eigenvalue we will refer to the fixed point as a spiral 
saddle~\cite{Sprott2003}. Fixed points with three real-valued negative 
eigenvalues of the Jacobian are referred to as stable nodes. The resulting 
phase diagram for the Moran dynamics is shown in 
Fig.~\ref{fig:IIPDE_M_heatmap}. The other three types of microscopic 
dynamics give qualitatively similar phase diagrams, but the exact 
quantitative positions of the various phase lines will generally be different. 

\begin{figure}
\vspace{2em}
\centering
\includegraphics[scale=0.3]{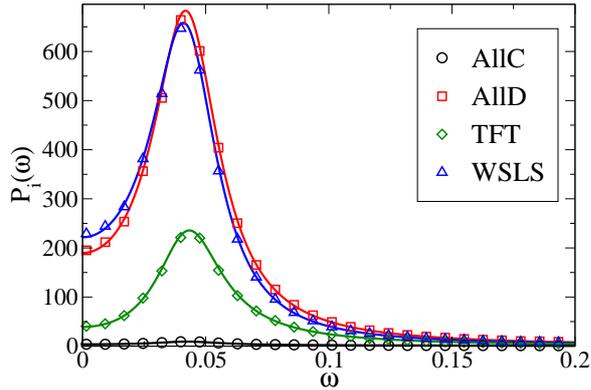}
\caption{(Color online) The power spectra for the four strategies with Moran 
updating at $u=0.02$ and $\epsilon=0.01$. Symbols are results from numerical 
simulations at $N=12000$, solid lines are the predictions of 
Eq.~(\ref{eqn:ps}).}
\label{fig:iipde_mps}
\end{figure}

When the mixed fixed point is a spiral saddle the deterministic dynamics 
can either converge to a limit cycle or to the attractor at AllD, depending 
on initial conditions. For locations in the parameter space where the mixed 
fixed point of the deterministic dynamics is a stable spiral, we again 
observe demographic oscillations, and they can be characterized analytically 
in a manner similar to that discussed in the previous section. We depict 
the resulting power spectra for the Moran process in Fig.~\ref{fig:iipde_mps}. 
As seen in the figure WSLS and AllD in particular undergo strong demographic 
oscillations, with a comparable magnitude between the two strategies.


\begin{figure}
\vspace{2em}
\centering
\includegraphics[scale=0.3]{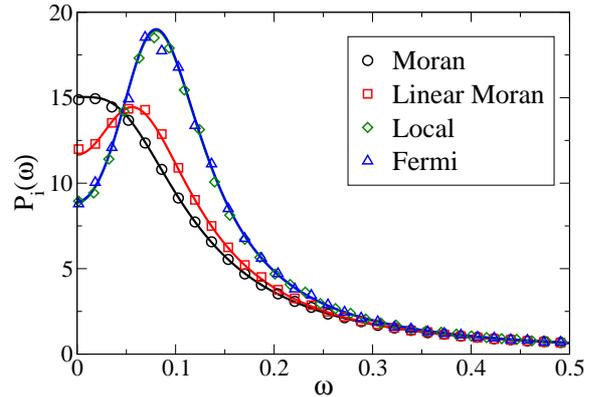}
\caption{(Color online) The oscillations in the TFT strategy concentration 
for the four update rules at $u=0.05$ and $\epsilon = 10^{-4}$. Symbols are 
from simulations, lines from the theory. System sizes in the range 
$N = 10^4$ to $10^6$ are used. For this particular choice of parameters, 
the spectra for the local and Fermi processes overlap, although this is 
not the case in general.}
\label{fig:IIPDE_ps}
\end{figure}

As with the iterated PD considered earlier, the amplitude of quasi-cycles 
resulting from an amplification of intrinsic fluctuations can be expected 
to be large when a fixed point of the stable spiral type is located close 
to the border between the stable-spiral and limit-cycle phases. There can 
then again be periods in time when WSLS is the most prevalent strategy, 
despite AllD dominating the fixed point. The power spectra for oscillations 
in the concentration of the TFT strategy resulting from the four different 
update rules are compared in Fig.~\ref{fig:IIPDE_ps}. We again observe 
that the Fermi process exhibits oscillations of a higher frequency and 
with a larger amplitude than the other rules. Although spectra for the 
local process and the Fermi process overlap in the figure, this is 
coincidental at this point in parameter space, and will not be the case 
in general.


\section{Conclusion and outlook}
\label{sec:conclusion}

In this paper we have used analytical approaches based on van Kampen's 
system-size expansion to study the emergence of quasi-cycles in 
evolutionary games in finite populations. Most existing studies of such 
effects are of a numerical nature; we have complemented these giving a 
systematic account of the formalism, and derived the resulting effective 
Langevin equations which describe the statistics and correlations of 
fluctuations in large, but finite populations. This approach and our general 
formulae are applicable to a large class of microscopic update rules, and
to games with an arbitrary number of pure strategies. They are, in principle, 
also valid for arbitrary mutation matrices. The results of this paper 
hence allow one to predict the regions in parameter space in which coherent 
quasi-cycles are to be expected, and to compute their spectral properties. 
In particular coherent cycles, such as reported in game dynamical systems 
e.g.~in~\cite{Imhof2005}, can be understood as a consequence of the 
combination of intrinsic noise and the existence of a stable fixed point with 
complex eigenvalues in the corresponding deterministic system obtained in 
the limit of infinite populations. In absence of noise the deterministic 
system approaches such fixed points in an oscillatory manner, with 
oscillations dying away at a rate proportional to the inverse real part
of the relevant eigenvalue. In finite systems, discretization noise leads 
to persistent stochastic corrections perturbing the system at all times. 
In the limit of large, but finite system sizes these fluctuations (the 
noise $\bm{\eta}$ in Eq.~(\ref{eqn:langevin}), together with the pre-factor 
$N^{-1/2}$) can be seen as a small perturbation to the deterministic 
dynamics, driving the system away from the fixed point. The attracting 
fixed point and the oscillatory approach to it on the deterministic level 
on the one hand and the persistent intrinsic noise on the other then 
conspire to give coherent and sustained stochastic cycles, with an 
amplitude largely determined by the inverse real part of the least stable 
complex eigenvalue.

We have applied the van Kampen formalism to the specific example of the
iterated PD game, where stochastic oscillations have been reported in the
earlier numerical study~\cite{Imhof2005}. We have worked out detailed 
phase diagrams depicting the nature of the limiting deterministic dynamics 
and we have studied systematically how the mutation rate and complexity 
cost, the two main model parameters, affect the outcome of the deterministic 
system. Based on this analysis we are able to predict the parameter regimes 
in which stochastic oscillations occur. In particular we find that 
oscillation amplitudes become maximal when the Hopf bifurcation line in 
the phase diagram is approached from within the stable phase. At the 
bifurcation line the oscillation amplitude formally diverges, with the 
power spectrum turning into a delta-function, and as the instability line 
is crossed a limit cycle is born. 

We have also carried out a detailed comparison of four different microscopic 
update rules; results indicate that their respective phase diagrams are 
qualitatively similar. The analysis shows that, at fixed values of the 
model parameters, the Fermi process tends to produce stochastic cycles 
with larger amplitudes and frequencies than the other update rules. We have 
extended our study to a version of the iterated PD game in which errors 
of the trembling-hand type occur with a small, but non-zero rate. The 
so-called win-stay lose-shift strategy has here been seen to out compete 
tit-for-tat, and accordingly we have considered a four-strategy space 
(always defect, always co-operate, tit-for-tat and win-stay lose-shift), 
and have identified the regions in parameter space where coherent cycles 
are most likely to occur. Analytical results for the resulting power spectra 
of these quasi-cycles are confirmed convincingly in numerical simulations.

Mathematical techniques of the type we have used here, most notably the 
master equation formalism and system-size expansions, were first devised 
in statistical physics, but they are becoming increasingly more popular 
in the game theory literature. This extends to equivalent approaches based 
on Kramers-Moyal expansions. We attribute this popularity to the generality 
with which these methods are applicable and to the fact that they allow 
one to obtain an exhaustive account of the properties of first-order 
stochastic corrections to the limiting deterministic dynamics. Exact 
analytical results can be derived for large, but finite populations, and 
hence these techniques make simulations on the microscopic level 
redundant (at least in principle). We expect this to be the case for 
games with more complicated strategy structures, or with interaction 
between more than two players such as for example in public goods games. 
The analytical approach can also be expected to be applicable to other, 
potentially more intricate types of human error. For such games it may 
be difficult or time consuming to carry out reliable simulations and 
to perform exhaustive parameter searches. Analytical characterizations 
of stochastic effects such as the ones discussed in this paper may 
then be particularly welcome.

\begin{acknowledgments} 
We would like to thank Sven van Segbroek for making his modification of 
the Dynamo package for Mathematica 6 available to us. TG would like to 
thank RCUK for support (RCUK reference EP/E500048/1). AJB acknowledges 
an EPSRC studentship.
\end{acknowledgments}

\appendix
\section{System-size expansion of the master equation}
\label{app:vankampen}
The van Kampen system-size expansion has been extensively discussed elsewhere,
together with explicit examples~\cite{McKane2005, Alonso2007, Kampen2007}, 
and so here we will only briefly summarize the general idea and give the 
final results of the calculation that are relevant for this paper.

The starting point is the substitution of the ansatz (\ref{eqn:decomp}) into
the master equation (\ref{eqn:mastertwo}) --- or its generalization to more 
than three strategies. This yields an expansion in powers of $N^{-1/2}$, after
a re-scaling of time by a factor of $N$. To leading order ($N^{-1/2}$) the
deterministic equation (\ref{eqn:genrepmut}) is obtained. To next-to-leading 
order ($N^{-1}$) the Fokker-Planck equation (\ref{eqn:fokkerplanck}) is found.
This is defined in terms of two quantities: $C_i = \sum_k J_{ik}\xi_k$, where
$J$ is the Jacobian of the dynamics given by Eq.~(\ref{eqn:genrepmut}), and $B$ 
a symmetric matrix. Since we are interested in fluctuations about stationary 
states, both $J_{ij}$ and $B_{ij}$ are time-independent. 

The Jacobian can be obtained in a straightforward fashion once the dynamics
(\ref{eqn:genrepmut}) is known. The elements of the matrix $B$ are found from
the $N^{-1}$ terms in the system-size expansion to be
\begin{equation}
B_{ij} = \left\{ 
\begin{array}{l l}
\sum_{k \neq i} \left[ T_{i \to k}(\bm{x}) + T_{k \to i}(\bm{x}) \right]\,,  & 
\quad \mbox{if $i=j$}\\ \\
- \left[ T_{i \to j}(\bm{x}) + T_{j \to i}(\bm{x}) \right]\,, & 
\quad \mbox{if $i\neq j$}\\ \end{array} \right..
\end{equation}
Therefore the deterministic and stochastic dynamics to the order we are working
are entirely determined by $T_{i \to j} (\bm{x})$. This can be found 
by making the substitutions $(n_i/N)\to x_i$, $f_i\to f_i^\infty$ and 
$\phi\to\phi^\infty$ in Eq.~(\ref{eqn:transitions}), to obtain
\be
T_{i\rightarrow j}(\bm{x}) = \sum_k x_k x_i g_{ki}(\bm{f}^{\infty})q_{kj}.
\label{eqn:transitions_deter}
\ee
This explicitly shows how to construct $T_{i\rightarrow j}(\bm{x})$, once 
the process (defined by $g_{ki}(\bm{f})$) and the mutation matrix ($q_{ij}$) 
have been given.

\section{Payoff matrix for a four strategy, infinitely repeated PD with 
trembling hand errors}
\label{app:wslspayoff}
When two players meet and play the PD game over multiple rounds their state in 
round $\ell$ is defined by their actions in that round, e.g. player $1$ 
co-operates, player $2$ defects. The actions of the pair of players then 
determines the payoff they each receive. The payoff matrix for the 
infinitely repeated PD game with trembling hand errors is constructed by 
considering the stationary distributions of the state of each pair of 
players, to first order in $\epsilon$, the probability of a `trembling hand' 
error occurring~\cite{Imhof2007}. It is given by
\begin{widetext}
\begin{equation}
\bordermatrix{& AllC & AllD & TFT & WSLS \cr
AllC & R - \epsilon(2R - S - T) & S + \epsilon(R+P-2S) & R -
\epsilon(3R-T-2S) & (R+S)/2 + (\epsilon/2)\alpha \cr
AllD & T - \epsilon(2T-R-P) & P + \epsilon(S+T-2P) & P + 
\epsilon(S+2T-3P) & (P+T)/2 - (\epsilon/2)\alpha \cr
TFT & R + \epsilon(2T + S - 3R) & P + \epsilon(T+2S-3P) & \gamma & \gamma \cr
WSLS & (R+T)/2 + (\epsilon/2)\beta & (P+S)/2 - (\epsilon/2)\beta & \gamma & 
R + \epsilon(T+2P+S-4R)\cr}.
\label{eqn:iipdematrix}
\end{equation}
where $\alpha = (T + P - R - S)$, $\beta =  (S + P - R - T)$ and 
$\gamma = (T+R+P+S)/4$.
\end{widetext}

\end{document}